\documentclass[aps,prb,reprint,superscriptaddress,amsmath,amssymb]{revtex4-2}
\usepackage{graphicx}
\usepackage{dcolumn}
\usepackage{bm}
\usepackage{color}

\begin{document}


\title{Competition between helimagnetic and ferroquadrupolar orderings in a monoaxial chiral magnet DyNi$_3$Ga$_{9}$ studied by resonant x-ray diffraction}

\author{Mitsuru Tsukagoshi}
\email[]{d210320@hiroshima-u.ac.jp}
\affiliation{Department of Quantum Matter, ADSE, Hiroshima University, Higashi-Hiroshima 739-8530, Japan}
\author{Takeshi Matsumura}
\affiliation{Department of Quantum Matter, ADSE, Hiroshima University, Higashi-Hiroshima 739-8530, Japan}
\author{Shinji Michimura}
\affiliation{Department of Physics, Faculty of Science, Saitama University, Saitama 338-8570, Japan}
\author{Toshiya Inami}
\affiliation{Synchrotron Radiation Research Center, National Institutes for Quantum Science and Technology, Sayo, Hyogo 679-5148, Japan}
\author{Shigeo Ohara}
\affiliation{Department of Physical Science and Engineering, Graduate School of Engineering, Nagoya Institute of Technology, Nagoya 466-8555, Japan}


\date{\today}

\begin{abstract}
Successive phase transitions in a rare-earth monoaxial chiral magnet DyNi$_3$Ga$_{9}$ have been investigated by resonant x-ray diffraction. 
Magnetic dipole and electric quadrupole degrees of freedom arising from the large angular moment of $J=15/2$, in combination with the symmetric and antisymmetric exchange interactions and the crystal field anisotropy, give rise to competing ordered phases. 
We show that the antiferromagnetically coupled Dy moments in the $ab$-plane form an incommensurate helimagnetic order with $\bm{q}\sim(0, 0, 0.43)$ just below $T_{\text{N}}=10$ K, which further exhibits successive first-order transitions to the commensurate helimagnetic order with $\bm{q}=(0,0,0.5)$ at $T_{\text{N}}^{\;\prime}=9.0$ K, and to the canted antiferromagnetic order with $\bm{q}=(0,0,0)$ at $T_{\text{N}}^{\;\prime\prime}=8.5$ K, both with large coexistence regions. 
The relation of the magnetic helicity and the crystal chirality in DyNi$_3$Ga$_{9}$ is also uniquely determined. 
Splitting of the $(6,0,0)$ Bragg peak is observed below $T_{\text{N}}^{\;\prime\prime}$, reflecting the lattice distortion due to the ferroquadrupole order. 
In the canted antiferromagnetic phase, a spin-flop transition takes place at 5 K when the temperature is swept in a weak magnetic field. 
We discuss these transitions from the viewpoint of competing energies described above. 
\end{abstract}

\maketitle

\section{Introduction}
\label{sec:I}
In chiral magnetic materials with crystal structures lacking both the space inversion and mirror symmetries, nontrivial magnetic structures can often be realized as a result of competing energies of symmetric magnetic exchange interaction, Dzyaloshinskii-Moriya (DM) antisymmetric exchange interaction, and the Zeeman energy by the external field. 

They are nonlinear topological textures of magnetic moments, which are typically exemplified by hexagonal crystallization of magnetic skyrmions in cubic B20-type $3d$ transition-metal compounds with the space group $P2_{1}3$~\cite{Muhlbauer09,Munzer10,Yu10,Adams12,Seki12}. 
Similar magnetic skyrmion phase is also realized in a $4f$-electron system EuPtSi, belonging to the same space group, with a much shorter helical pitch than in the $3d$ systems, providing an interesting contrast~\cite{Kaneko19,Tabata19,Sakakibara21,Hayami21}. 
In monoaxial chiral helimagnets such as CrNb$_3$S$_6$ (space group $P6_{3}22$), a periodic array of twisted spin structures is realized in magnetic fields applied perpendicular to the helical axis, which is called a chiral soliton lattice (CSL)~\cite{Togawa12,Togawa15,Honda20}. 
Underlying these phenomena is the spin-orbit coupling through which the electron spin sees the symmetry-broken space of the crystal~\cite{Kishine15,Inui20}. 

The $4f$ counterpart of CSL in CrNb$_3$S$_6$ is the Yb(Ni$_{1-x}$Cu$_{x}$)$_{3}$Al$_{9}$ system (space group $R32$)~\cite{Ohara14,Matsumura17a,Ninomiya18}. 
It was shown that the magnetic helicity and the crystal chirality has a one to one relation, indicating that they are coupled via the DM interaction~\cite{Matsumura17a}. 
The helimagnetic structures at zero field with incommensurate propagation vectors ranging from $\bm{q}=(0, 0, 0.82)$ for $x=0$ to $(0, 0, 0.44)$ for $x=0.06$, which is probably determined by the Ruderman-Kittel-Kasuya-Yosida (RKKY) exchange interaction, continuously transforms into the CSL state in magnetic fields applied perpendicular to the $c$-axis. 

One of the important backgrounds for the proper helical structure realized in Yb(Ni$_{1-x}$Cu$_{x}$)$_{3}$Al$_{9}$ is that the crystalline-electric-field (CEF) ground state is a well isolated magnetic doublet, allowing only the magnetic dipole degree of freedom and realizing the negligibly small magnetic anisotropy within the basal $ab$-plane. 
The first excited doublet is estimated to be at 47 K~\cite{Yamashita12}.
A similar aspect in EuPtSi is that it is a $S=7/2$ spin system without orbital moment, resulting in weak magnetic anisotropy. 
The weak anisotropy is considered to be an important background for the formation of nontrivial spin textures reflecting the intrinsic magnetic interactions. 

\begin{figure}[t]
\begin{center}
\includegraphics[width=8cm]{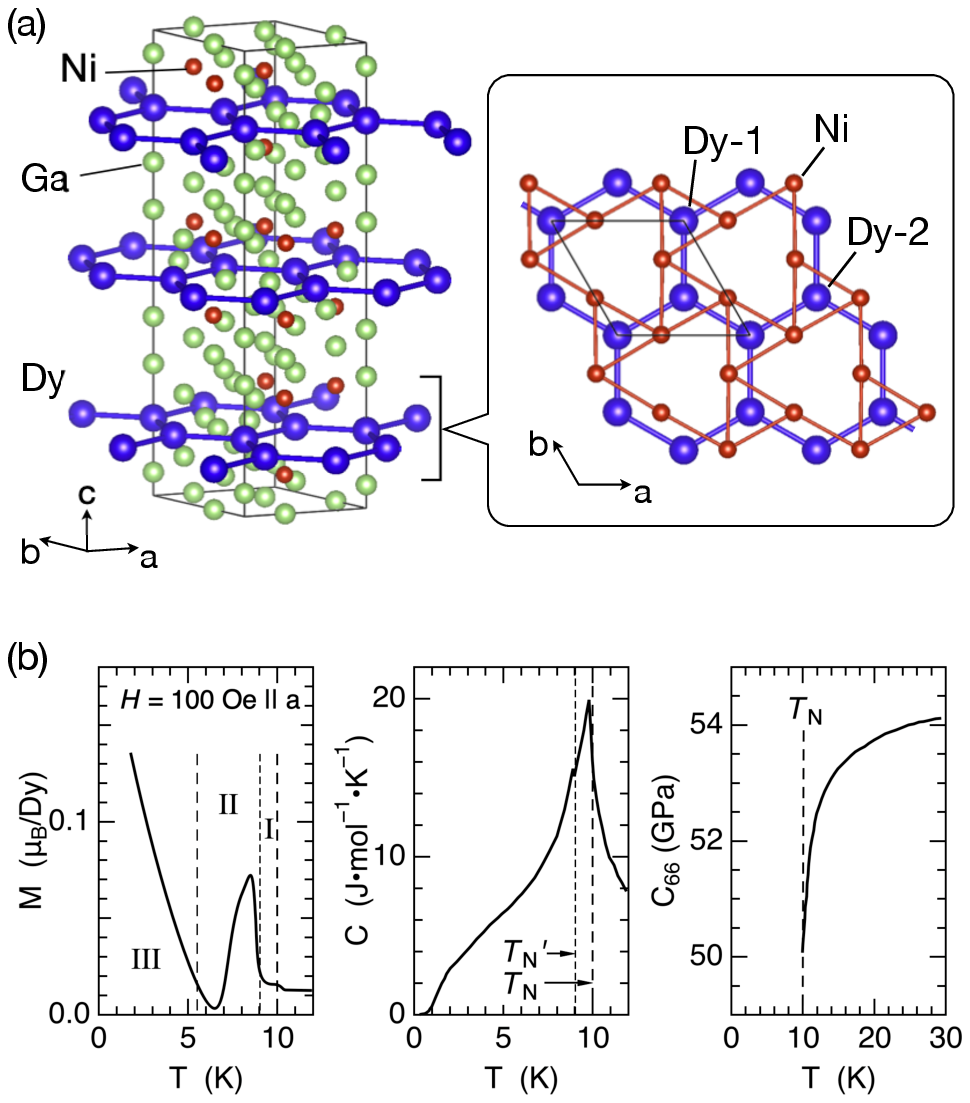}
\end{center}
\caption{(a) Crystal structure of DyNi$_3$Ga$_9$. The top view shows the Dy$_2$($6c$) + Ni$_6$($18f$) block layer at the bottom of the hexagonal unit cell, where the Ga atoms are omitted.  
The Dy atoms are located at the Wycoff positions $(0, 0, z)$ (Dy-1) and $(0, 0, -z)$ (Dy-2) on the rhombohedral lattice points $(0, 0, 0)$, $(2/3, 1/3, 1/3)$, and $(1/3, 2/3, 2/3)$, forming almost perfect honeycomb layers with $z=0.16697(3)$~\cite{Ninomiya17}.
VESTA was used to draw the figure~\cite{Momma11}. 
(b) Temperature dependences of magnetization at low field, specific heat, and $C_{66}$ elastic modulus of DyNi$_3$Ga$_9$ reproduced from the literatures~\cite{Ninomiya17,Ishii18}. 
}
\label{fig:CrystandProp}
\end{figure}

DyNi$_3$Ga$_9$, with the same crystal structure as YbNi$_3$Al$_9$ as shown in Fig.~\ref{fig:CrystandProp}(a), is a chiral magnet having contrasting features arising from the large angular moment of $J=15/2$ ($S=5/2$, $L=5$) of the Dy$^{3+}$ ion. 
The first and second excited doublet CEF-states are located at low energies around 10 K, giving rise to quadrupolar degrees of freedom in addition to the magnetic dipole moments to participate in the ordering phenomena. 
The detailed physical properties have been reported by Ninomiya \textit{et al.}~\cite{Ninomiya17}. 
DyNi$_3$Ga$_9$ exhibits successive phase transitions at $T_{\text{N}}=10$ K and $T_{\text{N}}^{\;\prime}=9$ K. 
The former transition at $T_{\text{N}}$ is considered to be a ferroquadrupole (FQ) order of $O_{xy}$ or $O_{22}$ because a huge and almost divergent elastic softening is observed in the $C_{66}$ mode whereas the magnetic susceptibility shows only a weak anomaly~\cite{Ishii18,Ishii19}. 
The latter one at $T_{\text{N}}^{\;\prime}$ is considered to be a canted antiferromagnetic (AFM) order with a ferromagnetic component, where the ordered moments lie in the $ab$-plane~\cite{Ninomiya17}. 
These features, which we study in detail in this paper, are summarized in Fig.~\ref{fig:CrystandProp}(b).  
Multistep magnetization processes are observed at high fields for $H \parallel c$-plane, suggesting a magnetic frustration in the honeycomb layers of Dy and an underlying contribution of quadrupolar interactions~\cite{Silva17,Mendonca18}. 

By neutron diffraction, the FQ ordered state below $T_{\text{N}}$ has been shown to be coexistent with an incommensurate magnetic order described by $\bm{q}=(0, 0, \sim 0.45)$ in the region $T_{\text{N}}^{\;\prime} < T < T_{\text{N}}$ (phase I), suggesting a helimagnetic order rotating within the $ab$-plane~\cite{Ninomiya17}. 
This is a controversial but interesting result in that the FQ order should confine the direction of the magnetic moments and prevent the helimagnetic order. There is also a possibility that a helical ordering of quadrupole moments coexists. 
Below $T_{\text{N}}^{\;\prime}$ (phase II), the magnetic structure changes to a commensurate one described by $\bm{q}=(0, 0, 0)$ and (0, 0, 0.5)~\cite{Ninomiya17}. 

Another interesting feature in DyNi$_3$Ga$_9$ is the phase II--III transition at 6 K, where the magnetic susceptibility vanishes~\cite{Ninomiya17}. 
Although this is seemingly reminiscent of a magnetic compensation in ferrimagnets, the true mechanism should be different and should be explained in association with the canted-AFM structure. 
To understand the roles of magnetic and quadrupolar exchange interactions, DM interaction, and the magnetic anisotropy on the successive transitions of DyNi$_3$Ga$_9$ described above, it is necessary to investigate the ordered structures of magnetic and quadrupolar moments from microscopic viewpoints, which is the purpose of this work. 

After describing the experimental procedure in Sec.~\ref{sec:II}, the results and analyses are presented in Sec.~\ref{sec:III}. 
In Sec.~\ref{sec:IIIA}, we show that the incommensurate helimagnetic order with $\bm{q}\sim(0, 0, 0.43)$ is realized just below $T_{\text{N}}$=10 K, followed by a first-order transition at $T_{\text{N}}^{\;\prime}$=9.0 K to the commensurate helimagnetic order with $\bm{q}=(0, 0, 0.5)$ and $(0, 0, 1.5)$. 
This again experiences a first-order transition at $T_{\text{N}}^{\;\prime\prime}$=8.5 K to the canted-AFM phase with $\bm{q}=(0, 0, 0)$.   
In Sec.~\ref{sec:IIIB}, the incommensurate and commensurate helimagnetic structures are analyzed and shown to have single and mixed helicities, respectively. A model structure is presented in Sec.~\ref{sec:IIIC}. 
In Sec.~\ref{sec:IIID}, we show that the phase II--III transition in the canted-AFM phase is a spin-flop transition by temperature. 
Evidence for a lattice distortion reflecting the FQ order is also presented. 
Based on these results, we discuss the successive phase transitions in Sec.~\ref{sec:IV}. 
After describing the consequences of the lattice distortion in Sec.~\ref{sec:IVA}, the mechanism of the spin-flop is discussed in Sec.~\ref{sec:IVB}. Finally, in Sec.~\ref{sec:IVC}, we discuss how the incommensurate and commensurate helimagnetic orderings are realized in DyNi$_3$Ga$_9$ from the viewpoint of competing energies of the DM interaction, RKKY interaction, and the CEF anisotropy.

\section{Experiment}
\label{sec:II}
\begin{figure}[t]
\begin{center}
\includegraphics[width=8.5cm]{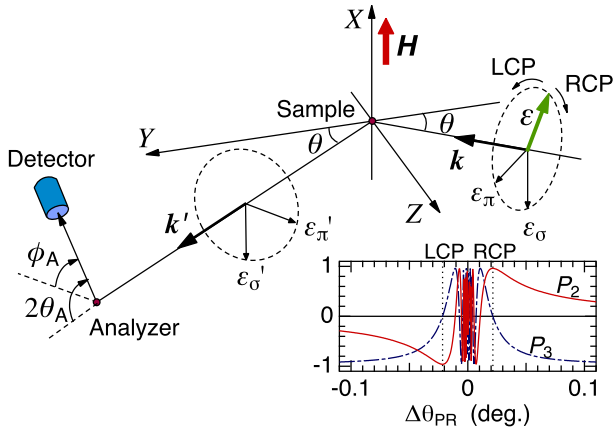}
\end{center}
\caption{Scattering configuration of the experiment. We take the $Z$-axis along the scattering vector $\bm{Q}=\bm{k}'-\bm{k}$, $X$-axis along  $\bm{k} \times \bm{k}'$, and $Y$-axis along $\bm{k} + \bm{k}'$. 
The inset figure shows the offset angle ($\Delta \theta_{\text{PR}}=\theta_{\text{PR}} - \theta_{\text{B}}$) dependence of the Stokes parameters $P_2$ and $P_3$, representing the polarization state after transmitting the phase plates inserted in the incident beam. 
$P_2$ and $P_3$ represent the degree of circular ($+1$ for RCP and $-1$ for LCP) and linear ($+1$ for $\sigma$ and $-1$ for $\pi$) polarizations, respectively. The vertical dotted lines stand for the positions of LCP and RCP states. 
The beam is depolarized in the region $\Delta \theta_{\text{PR}} \approx 0^{\circ}$. 
}
\label{fig:ScattConfig}
\end{figure}
Single crystals of DyNi$_3$Ga$_9$ were prepared by the Ga-flux method~\cite{Ninomiya17}. 
Resonant x-ray diffraction (RXD) experiments were performed at BL22XU of SPring-8 and at BL-3A of the Photon Factory, KEK, Japan. 
The samples were polished to a shining surface and were mounted in a vertical-field 8 T superconducting cryomagnet. Two samples were used, one with the $ab$-plane surface for the main RXD experiments in the $h0l$ scattering plane and the other with the (100) surface for nonresonant diffraction to investigate the lattice distortion by measuring the $h00$ reflection. 
The scattering geometry is shown in Fig.~\ref{fig:ScattConfig}. In normal conditions, the incident x ray from the synchrotron source is $\pi$-polarized with its electric field parallel to the scattering plane. 
The polarization of the diffracted x ray is analyzed using the 006 reflection of a pyrolytic-graphite crystal analyzer ($2\theta_{\text{A}}=90.7^{\circ}$ at 7.794 keV) before the x-ray photons are counted by a silicon drift detector (XR-100, Amptek). 

By inserting a diamond phase retarder system in the incident beam, we can tune the incident linear polarization to right-handed circular polarization (RCP) or left-handed circular polarization (LCP) by rotating the angle of the diamond phase-plate ($\theta_{\text{PR}}$) about the 220 Bragg angle ($\theta_{\text{B}}$), where the scattering plane is tilted by $45^{\circ}$ from the horizontal plane. 
By changing the offset angle $\Delta\theta_{\text{PR}} =  \theta_{\text{PR}} - \theta_{\text{B}}$, a phase difference arises between the $\sigma$ and $\pi$ components, which is approximately proportional to $1/(\theta_{\text{PR}} - \theta_{\text{B}})$. The polarization state of the incident beam as a function of $\Delta\theta_{\text{PR}}$ is shown in the inset using the Stokes parameters $P_2$ ($+1$ for RCP and $-1$ for LCP) and $P_3$ ($+1$ for $\sigma$ and $-1$ for $\pi$ linear polarization)~\cite{Lovesey96}. 
The polarization vectors of the RCP and LCP x rays are described by 
$\bm{\varepsilon}_+ = (\bm{\varepsilon}_{\sigma} + i \bm{\varepsilon}_{\pi}) e^{i(\bm{k}\cdot\bm{r} - \omega t)}$ and  
$\bm{\varepsilon}_- = (\bm{\varepsilon}_{\sigma} - i \bm{\varepsilon}_{\pi}) e^{i(\bm{k}\cdot\bm{r} - \omega t)}$, respectively. 
In the present experiment, we used two diamond phase plates with thickness of 0.5 mm each to compensate for chromatic aberration~\cite{Inami13,Matsumura14}.

The crystal chirality of the sample used in the RXD experiment was determined by measuring the energy dependence of the intensity of the $(1, 1, 18)$ and $(\bar{1}, \bar{1}, 18)$ fundamental Bragg reflections around the absorption edges of Dy and Ni, which is described in the Appendix. 
By comparison with the calculated intensities, it turned out to be right handed; i.e., the atomic position of Ni ($18f$ site) is given by $x=0.3335$, $y=0.0060$, and $z=0.08452$~\cite{Ninomiya17}, and not by its mirror reflection.

\section{Results and Analysis}
\label{sec:III}
\subsection{Helimagnetic orderings below $T_{\text{N}}$}
\label{sec:IIIA}
\subsubsection{Coexistent order parameters and the first order transitions}
\begin{figure}[t]
\begin{center}
\includegraphics[width=8cm]{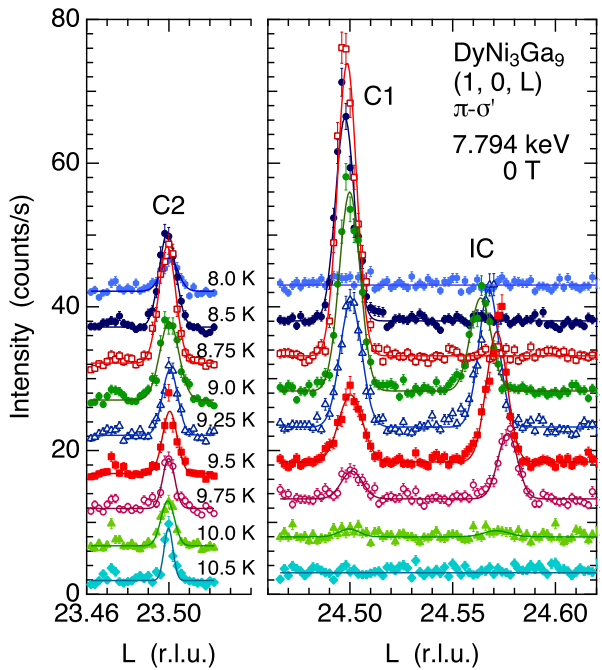}
\end{center}
\caption{Temperature dependence of the reciprocal space scan along $(1, 0, L)$ at zero field at the $E1$ resonance energy of 7.794 keV for the $\pi$-$\sigma'$ scattering channel. Solid lines are the fits with Gaussian functions. }
\label{fig:LscanPS}
\end{figure}

Figure~\ref{fig:LscanPS} shows the temperature ($T$) dependence of the reciprocal space scan along $(1, 0, L)$. 
The measurement was performed at zero field with increasing $T$ at the $E1$ resonance energy of 7.794 keV for the $\pi$-$\sigma'$ scattering channel. 
In phase I, three peaks are identified at an incommensurate wave vector labeled IC and at commensurate wave vectors labeled C1 and C2. 
The propagation vectors are expressed as $\bm{q}_{\text{IC}} = (0, 0, q)$ $(q\sim 0.43)$, $\bm{q}_{1} = (0, 0, 0.5)$, and $\bm{q}_{2} = (0, 0, 1.5)$~\cite{Note1}. 
The incommensurate wave vector clearly exhibits a $T$-dependence as in Yb(Ni$_{1-x}$Cu$_{x}$)$_{3}$Al$_{9}$. 
The C1 and C2 peaks appear above 8.5 K and vanish at 10 K, whereas the IC peak is observed above 9.0 K. 
Note that the C2 peak is superimposed on an extrinsic peak due to some scattering such as higher order reflections, which remains above 10 K and below 8 K. 
These peaks of IC, C1, and C2, probably of magnetic dipole origin, were also observed at other equivalent positions along $(-1, 0, L)$. 
However, none of them were detected along $(0, 0, L)$, indicating that the moments of Dy-1 and Dy-2 on the same layer are antiferromagnetically coupled. 
This is simply because the magnetic structure factor for the Bragg peak at $(0, 0, q)$ is described by $\sum_j (\bm{\mu}_{1,j}+\bm{\mu}_{2,j}) \exp (iqz_j)$, where $\bm{\mu}_{1,j}$ and $\bm{\mu}_{2,j}$ represents the magnetic moment of Dy-1 and Dy-2, respectively, on the $j$-th layer at $z=z_j$. This is not the case for the peaks at $(\pm 1, 0, q)$. 
\begin{figure}[t]
\begin{center}
\includegraphics[width=8cm]{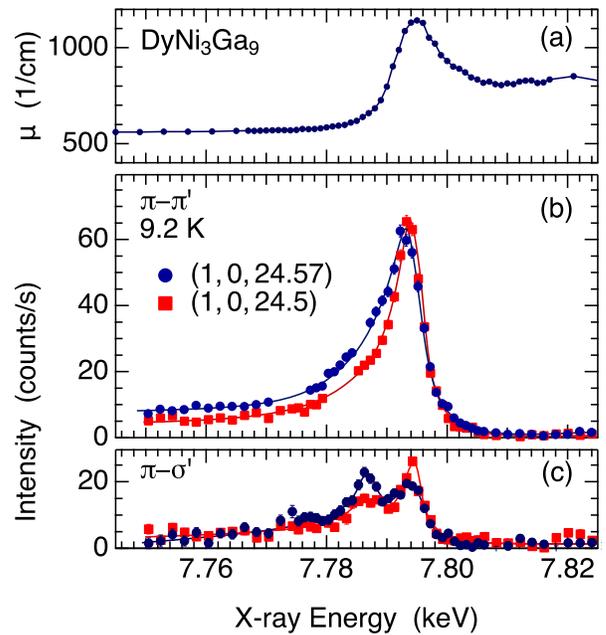}
\end{center}
\caption{(a) Energy dependence of the absorption coefficient obtained from the fluorescence spectrum.  
(b) and (c) Energy dependences of the (1, 0, 24.57) and (1, 0, 24.5) peak intensities at 9.2 K in phase I for the $\pi$-$\pi'$ and $\pi$-$\sigma'$ scattering channels, respectively. Solid lines are the guide for the eye. }
\label{fig:Escans}
\end{figure}

The energy dependences of the IC and C1 peaks are shown in Fig.~\ref{fig:Escans}. 
The resonance enhancements of the intensity at the absorption edge of Dy show that the Bragg peaks in Fig.~\ref{fig:LscanPS} originate from the orderings of the Dy moments. 
The main peak at 7.794 keV can be assigned to the $E1$ resonance $(2p \leftrightarrow 5d)$. 
The weak peak at 7.786 keV observed only in the $\pi$-$\sigma'$ channel may be assigned to the $E2$ resonance $(2p \leftrightarrow 4f)$. 
The long tail in intensity to the low energy side below 7.76 keV is connected to the nonresonant magnetic scattering. 
\begin{figure}[t]
\begin{center}
\includegraphics[width=7.5cm]{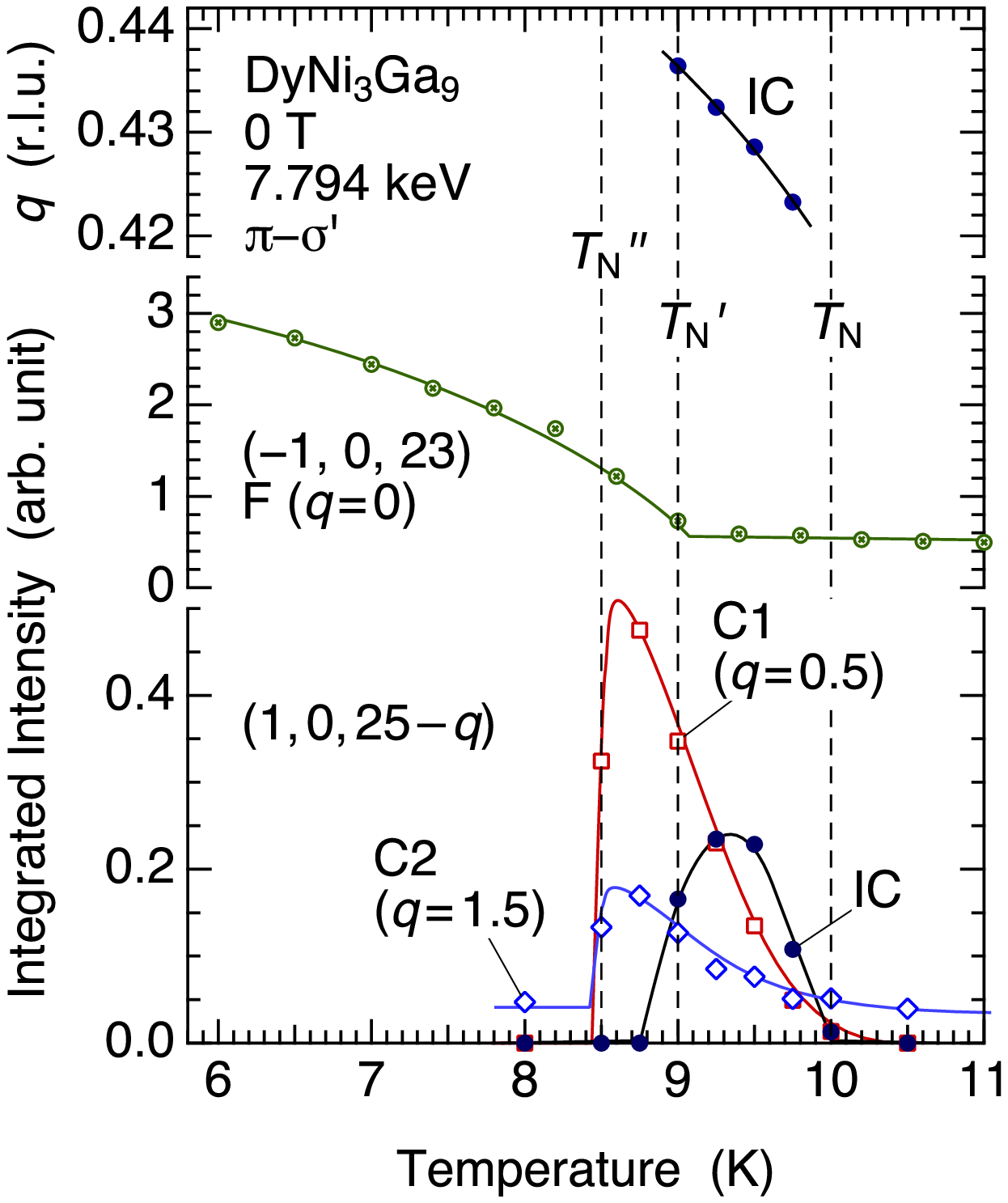}
\end{center}
\caption{Temperature dependence of the incommensurate $q$ value and the integrated intensities of the diffraction peaks at zero field for the $\pi$-$\sigma'$ scattering channel. Note that C2 has a constant background. 
}
\label{fig:TdepParams0T}
\end{figure}

Figure~\ref{fig:TdepParams0T} shows the $T$-dependence of the incommensurate $q$ value and the integrated intensities of the resonant Bragg peaks for $\pi$-$\sigma'$ shown in Fig.~\ref{fig:LscanPS}. The $\pi$-$\sigma'$ data measured at the $(-1, 0, 23)$ fundamental reflection labeled F ($q=0$), representing the canted-AFM structure with a ferromagnetic component, are also plotted. 
The F peak at $q=0$ observed in phase II disappears at $T_{\text{N}}^{\;\prime}$, which is consistent with the disappearance of the ferromagnetic component in $\chi(T)$~\cite{Ninomiya17}. 
It is noted, however, that the C1 and C2 peaks appear abruptly at 8.5 K, which we named $T_{\text{N}}^{\;\prime\prime}$, and coexist with the disappearing F peak. 
The IC peak starts to develop above 8.8 K, which do not coincide with $T_{\text{N}}^{\;\prime}$, and coexists with the C1 and C2 peaks up to $T_{\text{N}}$. 
Since the $T$-dependences of the C1 and C2 peaks are the same, these peaks reflect the Fourier components of the identical magnetic structure realized in the sample, i.e., the C2 peak is the third harmonic of the C1 peak.

These coexistent features of the order parameters show that the transitions at $T_{\text{N}}^{\;\prime\prime}$ and $T_{\text{N}}^{\;\prime}$ are of first order. 
The first order nature of these transitions are more clearly demonstrated by the hysteresis in the $T$-dependences measured with increasing and decreasing $T$, which is shown in the Appendix. 
In Fig.~\ref{fig:TdepParams0T}, although the C1+C2 and the IC peaks share a wide region of coexistence up to $T_{\text{N}}$, which seems rather confusing, we consider that they are separated by the first order transition at $T_{\text{N}}^{\;\prime} = 9$ K. 
When the C1+C2 peaks appear abruptly at 8.5 K, the IC peak does not exist. 
When the IC peak starts to develop, the C1+C2 peaks start to diminish. 
These behaviors indicate that the phase I between $T_{\text{N}}^{\;\prime}$ and $T_{\text{N}}$ is an incommensurate phase, which locks in to the nearby commensurate structure below $T_{\text{N}}^{\;\prime}$. 

\begin{figure}[t]
\begin{center}
\includegraphics[width=8cm]{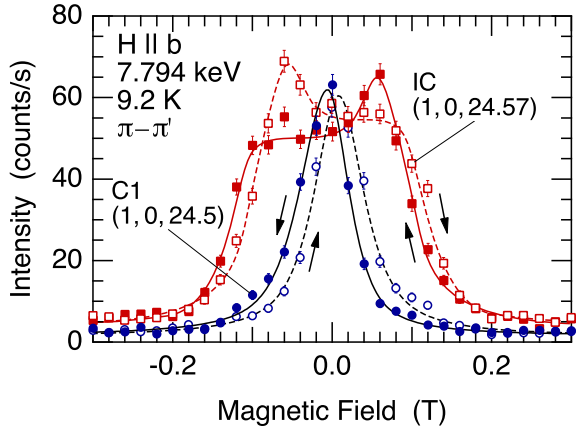}
\end{center}
\caption{Magnetic field dependence of the intensities of the incommensurate (IC) and commensurate (C1) peaks at 9.2 K for the $\pi$-$\pi'$ channel. The open and solid marks represent the field increase and decrease processes, respectively. }
\label{fig:HdepICC1}
\end{figure}

The change in the volume fraction between the IC and the C1+C2 phases can also be observed in the magnetic field dependence.  
Figure~\ref{fig:HdepICC1} shows the field dependence of the IC and the C1 intensities at 9.2 K in phase I measured in the $\pi$-$\pi'$ channel. 
The C1 and the IC peaks soon disappear by applying a magnetic field of $\sim 0.1$ T and $\sim 0.2$ T, respectively. 
However, the IC intensity increases when the C1 intensity decreases. 
This result also shows that the order parameter of phase I is incommensurate.

\subsubsection{Temperature dependent $q$ value of the IC structure}
Another noteworthy feature in Fig.~\ref{fig:TdepParams0T} is the $T$-dependent $q$ value of the IC peak. 
The  $q$ value $(\sim 0.43)$ is almost the same with that of Yb(Ni$_{1-x}$Cu$_{x}$)$_{3}$Al$_{9}$ for $x=0.06$~\cite{Matsumura17a}. 
In addition, the direction of the $T$-dependence of $q$, which decreases with increasing $T$, is also the same. 
It is noteworthy in DyNi$_3$Ga$_9$ that $q$ decreases from 0.44 to 0.425 in a narrow temperature range from 9 K to 10 K, which corresponds to 10 \% of $T_{\text{N}}$, whereas in Yb(Ni$_{1-x}$Cu$_{x}$)$_{3}$Al$_{9}$ for $x=0.06$ it requires $\sim$40 \% of $T_{\text{N}}$ to reach the same change.  This shows that the incommensurate $q$-vector of DyNi$_3$Ga$_9$ has almost four times stronger $T$-dependence than that of Yb(Ni$_{1-x}$Cu$_{x}$)$_{3}$Al$_{9}$ for $x=0.06$. This could be associated with the larger magnetic moment of Dy than that of Yb. 

The $T$-dependent $\bm{q}$-vector implies a modification in the RKKY-interaction $J(\bm{q})$ with the development of the ordered moment~\cite{JM91,Watson68,Elliot64}. 
The larger magnetic moment of Dy than that of Yb would give rise to a larger perturbation to the conduction electron system by the magnetic ordering, i.e, a larger modification in $\chi(\bm{q})$ and therefore in the exchange interaction $J(\bm{q})$ through the partial gap opening at the Fermi energy where $\varepsilon_{\bm{k}'}=\varepsilon_{\bm{k}+\bm{q}}$ is satisfied. As a result, the $T$-dependence of the $q$ value becomes almost proportional to the magnitude of the ordered moment. 
A similar $T$-dependent $\bm{q}$-vector has been reported mainly in isotropic Gd compounds~\cite{Feng13a,Feng13b,Detlefs96,Inami09,Matsumura17b,Matsumura19}. 
The present result in DyNi$_3$Ga$_9$ suggests that the Dy moments experience little in-plane anisotropy in phase I just below $T_{\text{N}}$. 
The weak anisotropy is considered to be a necessary condition for the formation of the helimagnetic structure.

\subsection{Analysis of the helimagnetic structure}
\label{sec:IIIB}
\subsubsection{Linear polarization analysis}
\begin{figure}[t]
\begin{center}
\includegraphics[width=8cm]{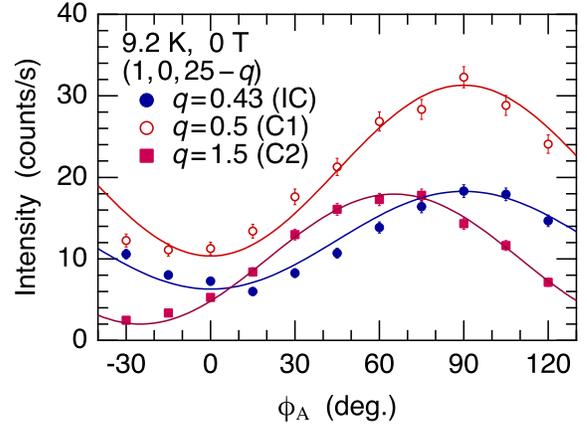}
\end{center}
\caption{Analyzer angle ($\phi_{\text{A}}$) dependences of the IC ($q=0.43$), C1 ($q=0.5$), and C2 ($q=1.5$) peak-intensities at 9.2 K for the $\pi$-polarized incident x ray. 
Solid lines are the calculations assuming a helical magnetic structure described in the text. }
\label{fig:POLICC1C2}
\end{figure}

The result of polarization analysis of the $(1, 0, 25-q)$ peaks at 9.2 K in phase I is shown in Fig.~\ref{fig:POLICC1C2}. 
With respect to the IC and C1 peaks, the intensity is minimum at $\phi_{\text{A}}=0^{\circ}$ and maximum at $90^{\circ}$. 
Since the background is approximately 1 cps, these data show that the intensity does not vanish at any angle. 
This result is well reproduced by assuming a proper helimagnetic structure with the moments lying in the $ab$-plane. 
The fact that no peak is detected along $(0, 0, L)$ indicates that the magnetic moments of Dy-1 and Dy-2 in the same honeycomb layer are oppositely directed. 
In the proper helical structure, the magnetic moments of Dy-1 and Dy-2 in the $i$-th layer at $z=z_i$ are expressed as 
\begin{align}
\bm{\mu}_{1,i} &=  \hat{\bm{x}}\; m_{qx} \cos q z_i + \hat{\bm{y}}\; m_{qy}  \cos (q z_i + \varphi)   \;, \\
\bm{\mu}_{2,i} &= - \bm{\mu}_{1,i} \;. 
\end{align}
For IC and C1 structures, we assumed $m_{qx}=m_{qy}$ and $\varphi=\pm \pi/2$, which is the irreducible representation in the space group $R32$. 
When $\varphi=\pi/2$ ($-\pi/2$), the magnetic moment rotates clockwise (counterclockwise) when propagating along the $c$-axis. 
Since we use the linearly polarized incident beam and analyze the linear polarization, the same calculated curves are obtained for $\varphi=\pm \pi/2$; 
although we cannot distinguish the magnetic helicity in this measurement, we can check if the proper helical model is appropriate or not. 
As demonstrated in Fig.~\ref{fig:POLICC1C2}, the calculated curves for IC and C1 well explain the data. 

The result of polarization analysis for the C2 peak is different. The intensity vanishes at $\sim -25^{\circ}$ and takes a maximum at $\sim 65^{\circ}$. 
This shows that the scattered x ray is linearly polarized. 
The calculated curve for C2 in Fig.~\ref{fig:POLICC1C2} is obtained by assuming a single domain state with $m_{qy}=-0.4 m_{qx}$ and $\varphi=0$, which is justified by the result that the intensity vanishes at $\phi_{\text{A}}=-25^{\circ}$.

\subsubsection{Helicities of the IC and the C1 helimagnetic structures}
\begin{figure}[t]
\begin{center}
\includegraphics[width=8cm]{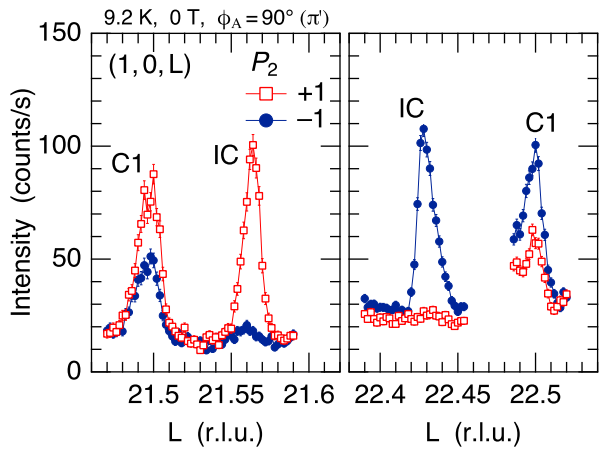}
\end{center}
\caption{Reciprocal space scan of the IC and C1 peaks at $(1, 0, 22 \pm q)$ for the RCP ($P_2=1$) and LCP ($P_2=-1$) x rays. 
Analyzer is set at $\phi_{\text{A}}=90^{\circ}$, detecting the $\pi'$ component. }
\label{fig:LscanCircICC1}
\end{figure}

To investigate the magnetic helicity of the IC and the C1 structures, we inserted a phase retarder system in the incident beam and measured the difference in intensity for RCP and LCP x rays. 
The result is shown in Fig.~\ref{fig:LscanCircICC1}. 
It is clearly demonstrated that the IC peak at $(1, 0, 22-q)$ is strong for RCP ($P_2=1$) and disappears for LCP ($P_2=-1$), whereas the relation is reversed at $(1, 0, 22+q)$. This shows that the magnetic helicity of the IC structure is uniquely determined. 
However, with respect to the C1 peak, the contrast between RCP and LCP incident x rays is not as perfect as that of the IC peak. 
Although the intensity at $(1, 0, 22-q)$ is stronger for RCP than for LCP, a finite intensity clearly exists for LCP. 

\begin{figure}[t]
\begin{center}
\includegraphics[width=8cm]{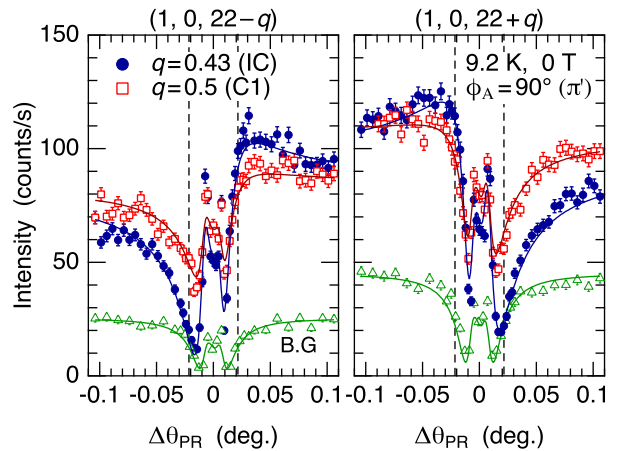}
\end{center}
\caption{Incident polarization ($\Delta \theta_{\text{PR}}$) dependence of the IC and C1 intensities at $(1, 0, 22 \pm q)$ with polarization analysis at $\phi_{\text{A}}=90^{\circ}$, detecting the $\pi'$ component. Vertical dashed lines represent the positions of circular polarization states. Triangles are the background. Solid lines are the calculations described in the text. }
\label{fig:PRscanICC1}
\end{figure}

Figure~\ref{fig:PRscanICC1} shows the incident polarization ($\Delta \theta_{\text{PR}}$) dependence of the IC and C1 intensities with polarization analysis at $\phi_{\text{A}}=90^{\circ}$. By scanning $\Delta \theta_{\text{PR}}$, the incident polarization changes as shown in Fig.~\ref{fig:ScattConfig}. 
As demonstrated in Fig.~\ref{fig:LscanCircICC1}, the $\Delta \theta_{\text{PR}}$ dependence is reversed for $(1, 0, 22 \pm q)$. 
The intensity of IC vanishes to the background level at $\Delta \theta_{\text{PR}}$ corresponding to the LCP and RCP positions for $(1, 0, 22-q)$ and $(1, 0, 22+q)$, respectively. However, the intensity for C1 does not vanish throughout the scan, although the asymmetric behavior is reversed. 
This is another piece of evidence with which we argue that the commensurate and the incommensurate phases are separated by the first order transition. 
If the identical magnetic structure had both the IC and C1 Fourier components, the C1 peak should also have a unique helicity, which is not the case. 
The calculated curves in Fig.~\ref{fig:PRscanICC1} were obtained by assuming a proper helimagnetic structure with a pure helicity of $\varphi=-\pi/2$ for the IC peak. 
To fit the data for the C1 peak, on the other hand, it is necessary to mix the intensity from the structure with opposite helicity ($\varphi=\pi/2$) by a ratio of 8 : 2. 

The calculated curves for IC and C1 in Figs.~\ref{fig:POLICC1C2} and \ref{fig:PRscanICC1} are obtained by assuming ideal helimagnetic structures with $m_{qx}=m_{qy}$ and $\varphi=-\pi/2$. 
Thus, we can conclude that the magnetic moments rotate counterclockwise when propagating along the $c$-axis in this sample, which turned out to be right handed as explained in the Appendix. 
Although the slight disagreements with the data can actually be improved by modifying these parameters, it is not the scope of the present work to discuss such a detailed modification from the ideally helical structure.

It was not possible to detect the $\sigma$-$\sigma'$ scattering for the IC and C1 peaks, which could have been crucial to identify the helical ordering of quadrupole moments by taking advantage of the fact that the magnetic scattering is forbidden in the $\sigma$-$\sigma'$ channel. 
We tuned the incident polarization to $\sigma$ using the phase retarder and scanned along $(1, 0, L)$ as in Fig.~\ref{fig:LscanPS}. 
However, there was a ridge of strong Thomson scattering along the $(1, 0, L)$ line above which we could not observe the resonant signal. 
This ridge of Thomson scattering arises from the stacking faults of the Dy$_2$Ni$_6$Ga$_{18}$ layers along the $c$-axis~\cite{Iba20}. 
This background signal is also observed in Figs.~\ref{fig:LscanCircICC1} and \ref{fig:PRscanICC1}. 
We assumed a Thomson scattering in fitting the background data in Fig.~\ref{fig:PRscanICC1}. 
The data for the IC and C1 were analyzed by treating the Thomson scattering to be included in the total scattering factor, where the magnetic and Thomson scatterings interfere, and not by treating the background as being superimposed on the magnetic scattering.

\begin{figure}[t]
\begin{center}
\includegraphics[width=7.5cm]{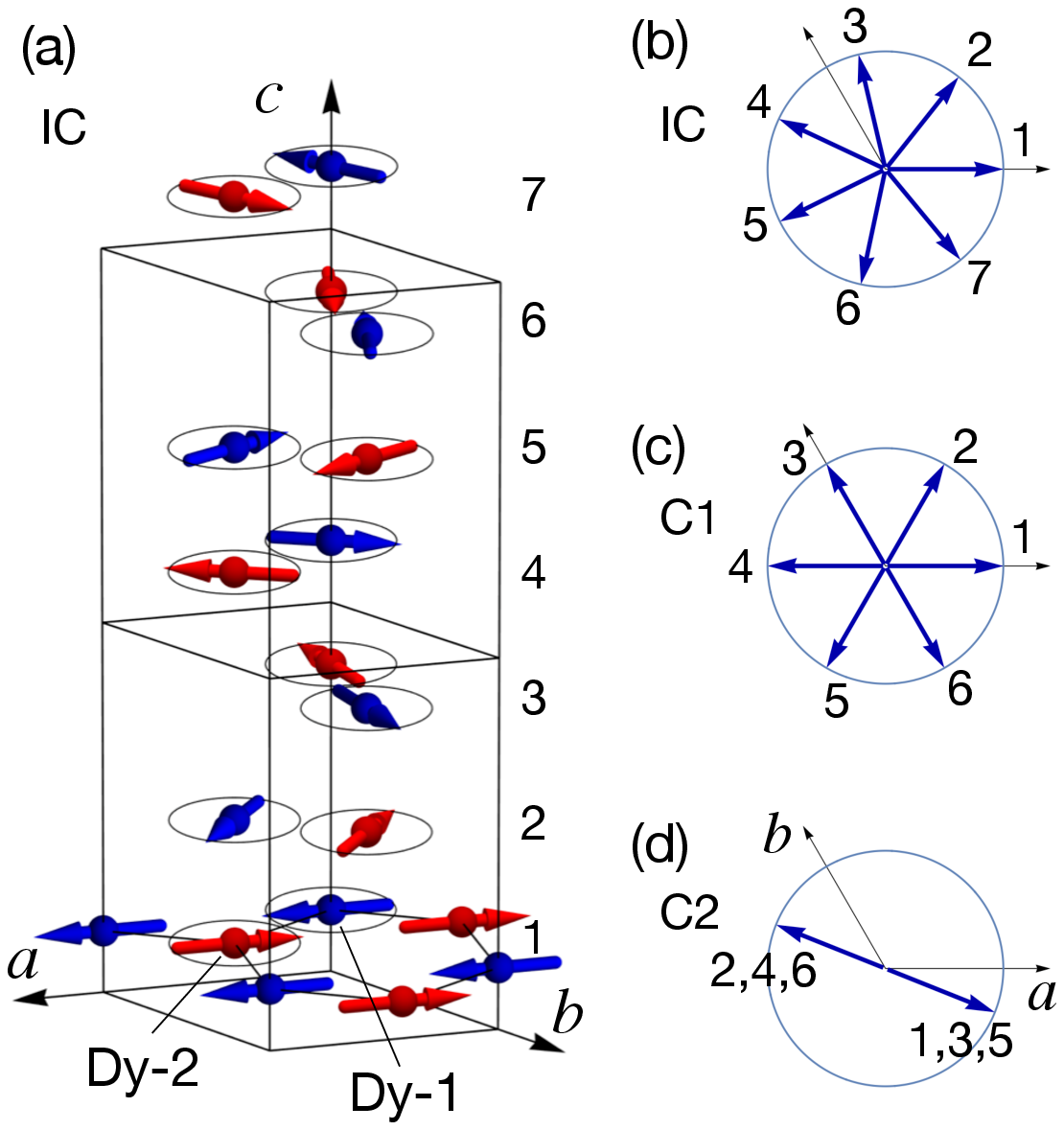}
\end{center}
\caption{(a) A model of the incommensurate helimagnetic structure with $q=0.43$. The magnetic moments on the Dy-1 and Dy-2 sites are colored by blue and red, respectively. (b) Directions of the magnetic moments of Dy-1, for $q=0.43$ (IC), on the seven honeycomb layers numbered in (a). 
(c) A model of the commensurate helimagnetic structure with $q=0.5$ (C1). 
(d) A model of the commensurate structure with $q=1.5$ (C2), which explains the data in Fig.~\ref{fig:POLICC1C2}.  }
\label{fig:Magst}
\end{figure}

\subsection{Helimagnetic structure model}
\label{sec:IIIC}
Models of the magnetic structures corresponding to the IC, C1, and C2 peaks are illustrated in Fig.~\ref{fig:Magst}. 
From the absence of the signal on the $(0, 0, L)$ line, it is concluded that the moments of Dy-1 and Dy-2 in the same honeycomb layer are antiferromagnetically coupled. 
In the IC phase between $T_{\text{N}}^{\;\prime}$ and $T_{\text{N}}$, the magnetic moment rotates counterclockwise when propagating along the $c$-axis in this right-handed crystal. 
It is noted that this relation of the magnetic helicity and the crystal chirality in DyNi$_3$Ga$_9$ is opposite to that in YbNi$_3$Al$_9$~\cite{Matsumura17a}.

The magnetic structure of Fig.~\ref{fig:Magst}(a) can be viewed as a combination of two interpenetrating rhombohedral sublattices of Dy-1 and Dy-2 moments, which are ferromagnetically ordered in a layer, antiferromagnetically coupled with each other, and helimagnetically ordered along the $c$-axis. 
Since the AFM coupling between Dy-1 and Dy-2 is maintained in the whole temperature range down to the lowest temperature, the intersublattice nearest-neighbor AFM interaction within a layer is considered to be the dominant exchange interaction. 
The helical pitch with $q\sim 0.43$ is determined probably by the much weaker RKKY interaction between the layers separated along the $c$-axis. 
The magnetic helicity is finally determined by the DM interaction, which is expected to be the weakest.  
The single helicity of the helimagnetic structure, however, indicates an important role of the DM antisymmetric interaction even in the presence of much stronger in-plane AFM interaction. 

In this incommensurate helimagnetic structure, the direction of the magnetic moments cover all the angles in the $ab$-plane, indicating that the IC phase is not influenced by the in-plane anisotropy. 
Although it becomes commensurate with the lattice accidentally when $q=3/7=0.4286$, there seems to be no indication of such a lock-in behavior since the $q$ value changes continuously with temperature. 

In the commensurate phase below $T_{\text{N}}^{\;\prime}$, the magnetic structure is described by a superposition of the C1 and C2 components. 
An example of the C1 and C2 structures are shown in Figs.~\ref{fig:Magst}(c) and \ref{fig:Magst}(d), respectively. 
The angle between the magnetic moments of the neighboring layers is exactly $60^{\circ}$ for C1 and $180^{\circ}$ for C2. 
It is noted that, although the C1 structure is illustrated in Fig.~\ref{fig:Magst}(c) as if the moments coincide with the $a$ axis, we cannot determine the direction of the moments in the $ab$-plane in the present set of the data. Fig.~\ref{fig:Magst}(c) is an example to show the six-fold helical structure, which is compatible with the hexagonal CEF anisotropy in the paramagnetic phase. 
The model structure of Fig.~\ref{fig:Magst}(d) for the C2 component, where the single-domain state is assumed, explains the data in Fig.~\ref{fig:POLICC1C2}. 
Although the actual magnetic structure of the C1+C2 phase should be described by a superposition of the C1 and C2 components, we do not have enough information on the phase relation between the two components.

\subsection{Canted-AFM order with $\bm{q}=(0,0,0)$ }
\label{sec:IIID}
\subsubsection{Spin-flop transition by temperature}
\begin{figure}[t]
\begin{center}
\includegraphics[width=7.5cm]{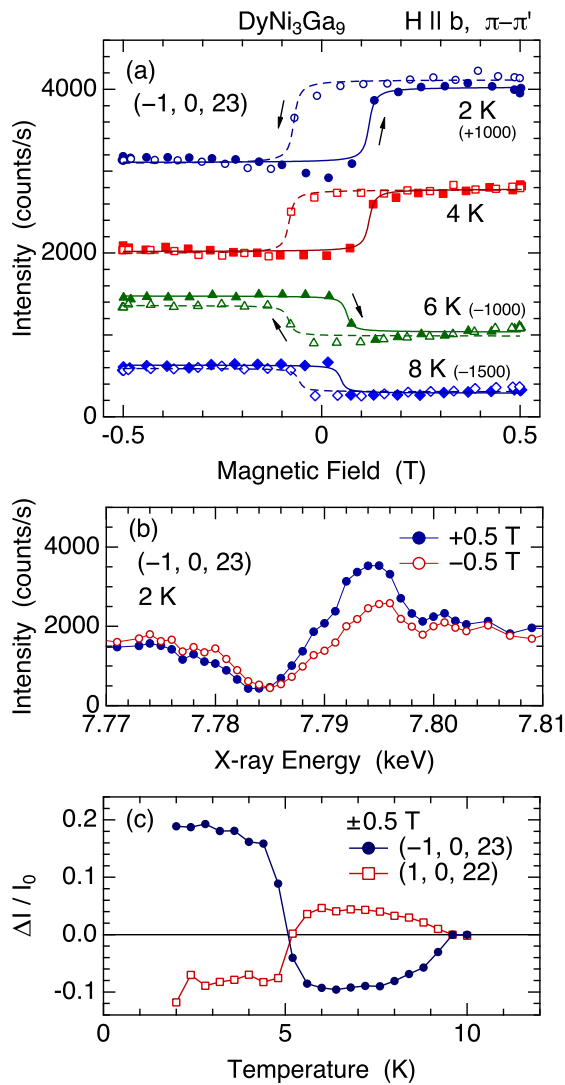}
\end{center}
\caption{(a) Magnetic field dependences of the intensity at the $E1$ resonance energy of 7.793 keV at four representative temperatures. 
Closed and open marks represent the field increase and decrease processes, respectively. Solid and dashed lines are the fits with step functions described in the text. 
(b) X-ray energy dependence of the $(-1, 0, 23)$ reflection $(q=0)$ for $H \parallel b$ at $\pm 0.5$ T in the $\pi$-$\pi'$ channel. 
(c) Temperature dependence of the intensity step ($\Delta I$) divided by the average intensity ($I_0$) for $(-1, 0, 23)$ and $(1, 0, 22)$. }
\label{fig:HdepFM}
\end{figure}

In phase II below the first order transition at $T_{\text{N}}^{\;\prime\prime}$, the the main magnetic structure is described by the propagation vector $\bm{q}=(0, 0, 0)$, reflecting the appearance of the canted-AFM structure. 
Although there remain a weak signal at $\bm{q}=(0, 0, 0.5)$ reflecting a slight modification of the canted-AFM structure~\cite{Ninomiya17}, which is also observed in our study, it is not a scope of this work to go deep into the details; it is much weaker than the C1 peak above $T_{\text{N}}^{\;\prime\prime}$. 
By applying a weak magnetic field in phase I, the IC and the C1 peaks soon disappear as shown in Fig.~\ref{fig:HdepICC1}, suggesting that the helimagnetic structure easily changes to the canted-AFM structure by a weak magnetic field of $0.1\sim 0.2$ T. 
Therefore, the main order parameter below $T_{\text{N}}^{\;\prime\prime}$ or in a weak magnetic field below $T_{\text{N}}$ is the canted-AFM. 

Figure~\ref{fig:HdepFM}(a) shows the magnetic field dependence of the intensity of the $(\bar{1}, 0, 23)$ fundamental reflection measured with increasing $T$ in the $\pi$-$\pi'$ channel at the $E1$ resonance energy. 
This measurement well captures the behavior of the canted-AFM moments. 
Initially at 2 K, the intensity steps up when the field is increased from negative to positive, whereas above 5 K, this behavior is reversed. 
This change in intensity when the field direction is reversed arises from the interference between the Thomson scattering and the resonant magnetic scattering as shown by the different energy spectrum in Fig.~\ref{fig:HdepFM}(b) for $\pm 0.5$ T. 
When the magnetic moments change their directions by the field reversal, the resonant magnetic structure factor $F_{M}$ changes its sign, whereas the crystal structure factor $F_{C}$ does not change. 
Since the observed intensity is proportional to $|F_{C} + \alpha(\omega) F_{M}|^2$, where $\alpha(\omega)$ represents a spectral function, the intensity exhibits a step up or down when the ferromagnetic moment changes its direction. 
The data in Fig.~\ref{fig:HdepFM}(a) are fit with a step function
\begin{equation}
 I = I_0 + \frac{2 \Delta I}{\pi} \arctan \frac{H-H_0}{\Gamma} \,,
\end{equation}
where $I_0$ represents the intensity of the fundamental reflection proportional to $|F_{C}|^2$
and $\Delta I$ the interference term proportional to $2F_{C}F_{M}$. 
The intensity ratio $\Delta I/I_0$ is plotted in Fig.~\ref{fig:HdepFM}(c). 
For the $(1, 0, 22)$ reflection, the sign of $\Delta I$ is reversed because the sign of $F_{C}$ is opposite to that of the $(\bar{1}, 0, 23)$ reflection. 
What is noteworthy with this result is that $F_{M}$ changes its sign at 5 K, which coincides with the temperature where the magnetic susceptibility vanishes~\cite{Ninomiya17}. 
This result indicates that some kind of spin-flop transition takes place by changing the temperature, which will be discussed in the next section.

\subsubsection{Lattice distortion below $T_{\text{N}}$ and $T_{\text{N}}^{\;\prime\prime}$}
\begin{figure}[t]
\begin{center}
\includegraphics[width=7.5cm]{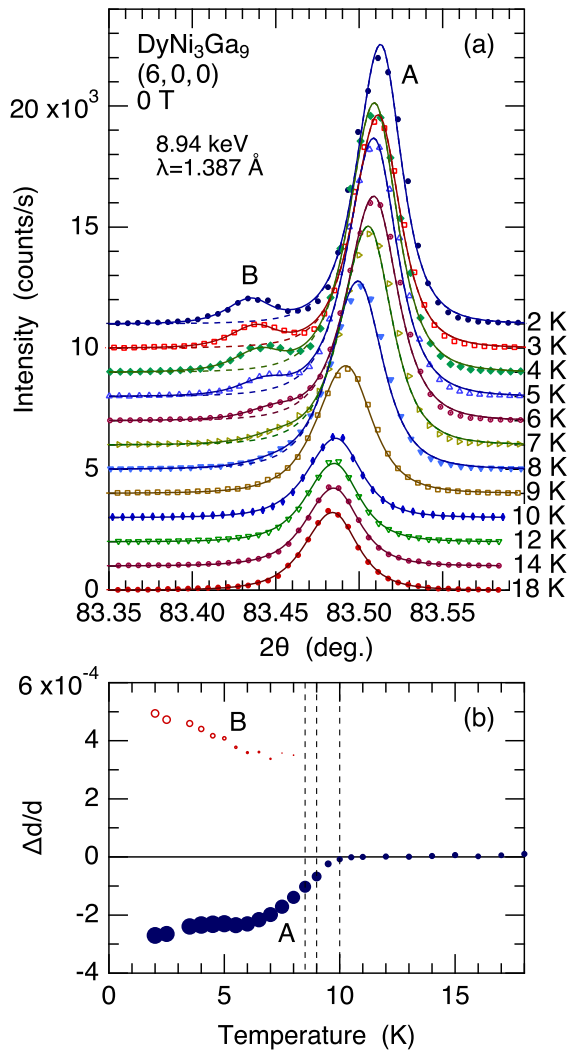}
\end{center}
\caption{(a) Temperature dependence of the peak profile of the $(6,0,0)$ fundamental Bragg reflection at zero field. Solid lines are the fits with asymmetric squared Lorentzian functions. Dashed lines represent the tails of the main peak A. 
(b) Temperature dependence of the relative change in the planar spacing. The size of the marks represents the intensity of the peak. 
The vertical lines represent the phase boundaries at $T_{\text{N}}^{\;\prime\prime}$, $T_{\text{N}}^{\;\prime}$, and $T_{\text{N}}$ shown in Fig.~\ref{fig:TdepParams0T}. 
}
\label{fig:Tdep600tth}
\end{figure}

Since $T_{\text{N}}$ has been considered to be a FQ order from the huge elastic softening in the $C_{66}$ mode, a lattice distortion is expected to occur in the $ab$-plane, reflecting the symmetry lowering from the hexagonal lattice. 
Figure~\ref{fig:Tdep600tth}(a) shows the $T$-dependence of the peak profile of the $(6,0,0)$ fundamental reflection measured with increasing $T$. 
At the lowest temperature of 2 K, the $(6,0,0)$ peak exhibits a clear split into two peaks, which are named A and B. 
The positions of the two peaks change with increasing $T$ and finally merge together above 10 K. The peak profiles have been fit with squared Lorentzian functions and the parameters obtained are summarized in Fig.~\ref{fig:Tdep600tth}(b) as the $T$-dependence of $\Delta d / d$ with its mark size representing the intensity. 
No splitting and no change in the peak position was observed in the $(0, 0, 27)$ reflection within the present experimental accuracy, indicating that the $d$-spacing along the $c$-axis does not change ($\Delta d/d < 1 \times 10^{-5}$). 

Below $T_{\text{N}}=10$ K, $\Delta d / d$ of the main peak A starts to decrease, which is associated with the helimagnetic order in the $ab$-plane. 
However, the peak splitting is not visible down to 8 K. 
Experimentally, it is not clear whether this is because the splitting is too small to be resolved, i.e., it exists and continuously increases from zero below $T_{\text{N}}$, or this is the intrinsic nature of the phase I, i.e., the hexagonal lattice is maintained with its $d$-spacing decreased. 
The latter possibility is probable if we remind that the elastic $C_{66}$ mode exhibits a strong attenuation in phase I, which can be associated with an unstable situation of the lattice which is close to lower its symmetry but still keeps the hexagonal lattice. 

The peak splitting is visible below 8 K, which seems to coincide with $T_{\text{N}}^{\;\prime\prime}$ where the helimagnetic structure vanishes and the canted-AFM structure with $q=0$ is stabilized. This is a direct piece of evidence of the symmetry lowering due to the FQ order. 
Another point to be noted is that the intensity of the main peak A is much stronger than the satellite peak B. 
If the lattice distortion takes place in the $ab$-plane, it should ideally result in the appearance of three domains, resulting in the intensity ratio of 2 : 1. 
This large disproportionation in the domain ratio could be due to a surface-strain effect. 
With respect to the increase in intensity of the main peak A below $T_{\text{N}}=10$ K, it is probably due to the reduction of the extinction effect by the occurrence of microscopic lattice distortion~\cite{SM1}.

\section{Discussions}
\label{sec:IV}
\subsection{Lattice distortion}
\label{sec:IVA}
The peak splitting of the $(6,0,0)$ reflection shows that there arise two different lattice parameters due to structural domains with a symmetry lower than trigonal, which is limited to the monoclinic lattice.  
The space group is most likely to be $C2$, since it is the only maximal monoclinic subgroup of $R32$.  
Although there remains a possibility of even lower triclinic symmetry, we restrict our discussion here within $C2$ since the basic idea does not change and also because it is beyond the accuracy of our data to discuss the triclinic structure. 

\begin{figure}[t]
\begin{center}
\includegraphics[width=8.5cm]{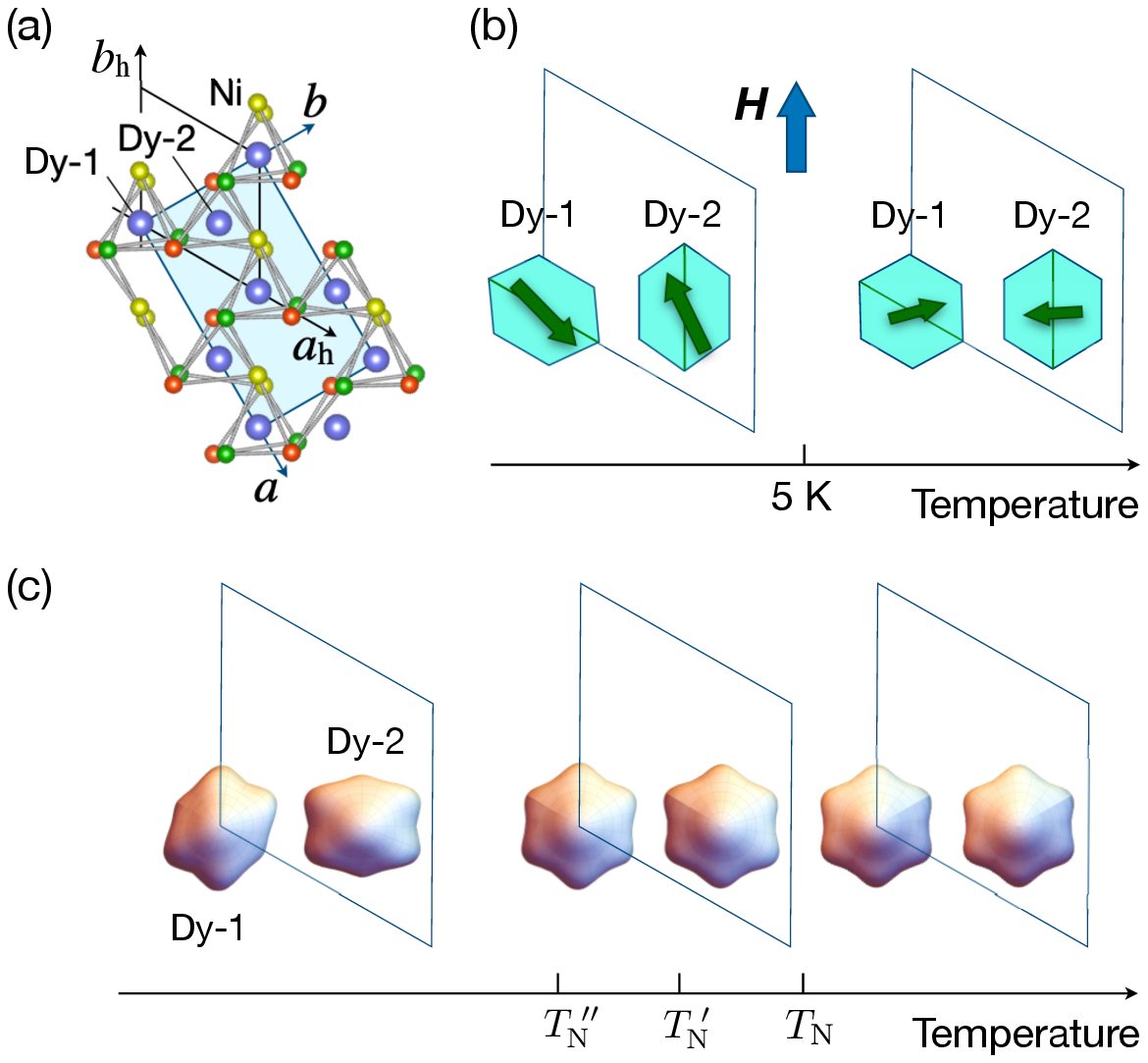}
\end{center}
\caption{(a) A possible model of atomic displacements of Ni in the monoclinic phase. 
The axes of the monoclinic unit cell are given by $\bm{a}=\bm{a}_{\text{h}}-\bm{b}_{\text{h}}$, $\bm{b}=\bm{a}_{\text{h}}+\bm{b}_{\text{h}}$ and $\bm{c}=(-\bm{a}_{\text{h}}+\bm{b}_{\text{h}}+\bm{c}_{\text{h}})/3$, where $\bm{a}_{\text{h}}$, $\bm{b}_{\text{h}}$, and $\bm{c}_{\text{h}}$ are the axes of the hexagonal unit cell. 
Only the nearest neighbor Ni atoms around the Dy atoms in a honeycomb layer at the bottom of the hexagonal unit cell are shown. 
The three Ni sites ($4c$) are represented by different colors, red, green, and yellow. The basal rectangular $ab$-plane of the monoclinic unit cell is colored. 
(b) A schematic of the antiferromagnetic structure with $q=0$ below $T_{\text{N}}$ in a magnetic field, exhibiting a spin-flop transition at 5 K. The hexagons represent the hexagonal CEF modified by the lattice distortion. The diagonal line represents the principal axis of the CEF. 
(c) Conceptual illustration of the charge distribution of Dy-$4f$ electrons in the paramagnetic phase, intermediate phase with the helimagnetic orderings, and the ferroquadrupole ordered phase. 
}
\label{fig:SpinFlop}
\end{figure}

In $C2$, all the atoms are located at the general $4c$ site. The three-fold symmetry at the Dy site is lost. 
The general $18f$ site of Ni and Ga in $R32$ is divided into three $4c$ sites. The unit-cell volume is 2/3 of that of $R32$. 
In Fig.~\ref{fig:SpinFlop}(a), we show, as an example, a simplified view of a model structure to represent the change in symmetry around Dy due to the displacements of Ni atoms. 
There are two other domains in which the $a$ and $b$ axes are rotated by $\pm 120^{\circ}$ from those of Fig.~\ref{fig:SpinFlop}(a), which allows two different $d$-spacings for the $(6,0,0)$ reflection in the hexagonal index. 

An important outcome of this symmetry lowering at the Dy site is the appearance of the orthorhombic CEF terms, i.e., $B_{22}$, $B_{42}$, $B_{44}$, $B_{62}$, and $B_{64}$, which do not exist in the trigonal structure.  
In addition, the principal axes of the CEF in the $ab$-plane are different between Dy-1 and Dy-2, resulting in different in-plane magnetic anisotropy as schematically represented by the hexagons shown in Fig.~\ref{fig:SpinFlop}(b). 
From the viewpoint of electric quadrupole, different linear combinations of $O_{22}$ and $O_{xy}$, which belong to the same irreducible representation in the hexagonal point group, are induced at Dy-1 and Dy-2, leaving finite quadrupole moment in total.
This state can be viewed as a FQ order with $\bm{q}=(0,0,0)$. A conceptual illustration of the ordered state is shown in Fig.~\ref{fig:SpinFlop}(c). 
Since the application of a shear strain in the $ab$-plane also induces such a situation, the large elastic softening in the $C_{66}$ mode is consistent with the spontaneous monoclinic distortion.

\subsection{Phase II-III spin-flop transition }
\label{sec:IVB}
The difference in the in-plane anisotropy between Dy-1 and Dy-2 causes the phase II--III transition at 5 K. 
At temperatures just below $T_{\text{N}}^{\;\prime\prime}$, the lattice distortion is still small as is observed by the weak intensity of the peak B in Fig.~\ref{fig:Tdep600tth}. 
The in-plane magnetic anisotropy is expected to be weak and the ordered quadrupole moments do not play important roles. 
When a magnetic field is applied in such a situation, the AFM moments prefer to be perpendicular to the applied field. 
With decreasing $T$, however, the magnitude of the ordered AFM moments as well as the lattice distortion increases, and thereby the modified CEF anisotropy would increase. 

Let us estimate the canted-AFM structure. 
Using the atomic sites reported in Ref.~\onlinecite{Ninomiya17}, we have the structure factors of 
$F_{(1,0,22)}=-2.76 f_{\text{Dy}} + 3.59 f_{\text{Ga}} -0.144 f_{\text{Ni}}$ and 
$F_{(\bar{1},0,23)}=3.25 f_{\text{Dy}} - 3.50 f_{\text{Ga}} +0.288 f_{\text{Ni}}$, which have opposite signs. 
First, we assumed the magnetic moments of Dy-1 and Dy-2 as written by $\bm{m}_1=(\cos \theta_1, \sin \theta_1, 0)$ and $\bm{m}_2=(\cos \theta_2, \sin \theta_2, 0)$. 
The relative angle between $\theta_1$ and $\theta_2$ was fixed so that the ferromagnetic moment $(\bm{m}_1 + \bm{m}_2)/2$ becomes 0.1 as observed in the $M(H)$ curve (10 \% of the saturation moment). 
Second, we assumed $\theta_1$ and $\theta_2$ are rotated by $180^{\circ}$ when the field direction is reversed, which leads to the sign change in $F_{\text{M}}$. 
Then, we searched for solutions in which $|F_{\text{C}} \pm \alpha F_{\text{M}}|^2$ roughly reproduces the intensity difference as observed in Fig.~\ref{fig:HdepFM}. We show one of the plausible solutions in Fig.~\ref{fig:SpinFlop}(b). 
Above 5 K, $\bm{m}_1$ and $\bm{m}_2$ are almost perpendicular to the field, but $\bm{m}_1$ is more tilted to the field direction than $\bm{m}_2$.  
In this temperature region, the Zeeman energy is more important than the anisotropy energy. 
Below 5 K, $\bm{m}_1$ and $\bm{m}_2$ are more influenced by the anisotropy. 
Then, $\bm{m}_2$ is more tilted to the field direction and $\bm{m}_1$ is directed opposite to the field, resulting in the sign reversal of $F_{\text{M}}$. 
We consider that this is the mechanism of the spin-flop transition by temperature. 
It is noted that Fig.~\ref{fig:SpinFlop}(b) is drawn so that $\bm{m}_1 \times \bm{m}_2 \parallel \bm{c}_{\text{h}}$ is satisfied to be consistent with the sense of the helimagnetic structure in phase I reflecting the DM interaction. 

\subsection{Helimagnetic orderings below $T_{\text{N}}$}
\label{sec:IVC}
Here we finally present our answer to the initial question. 
Since the helimagnetic order in DyNi$_3$Ga$_9$ is contradictory to the FQ order, it is allowed only in the temperature region just below $T_{\text{N}}$ where both the ordered moment and the lattice distortion are small. 
This feature is reflected in the $T$-dependence of the IC and the C1 peaks in Fig.~\ref{fig:TdepParams0T}. 
With decreasing $T$, the IC peak first develops just below $T_{\text{N}}$, indicating that the RKKY and the DM exchange interactions govern the magnetic order. 
The incommensurate $q$ value of $\sim 0.43$ reflects the maximum of the magnetic exchange interaction $J(\bm{q})$ along the $c$-axis, which is determined by the long range RKKY interaction. 
The chiral degeneracy of the helimagnetic structure is lifted by the weak DM interaction, which is uniquely determined by the crystal structure and allows only one helicity.  
The in-plane AFM interaction is expected to be much stronger than these interactions. 
The present result shows that the strongly coupled AFM moments in a layer form a helimagnetic rotation with a well defined single helicity due to the much weaker DM interaction. 
This makes an interesting contrast to the helimagnetic ordering in CrNb$_3$S$_6$, where the ferromagnetic ordered state due to strong in-plane and weak inter-layer ferromagnetic interaction is modified by the weak DM interaction to form the helimagnetic order~\cite{Shinozaki16}. 

The lattice distortion in this region, if existed, should be still very small, or could even be fluctuating without the symmetry lowering, i.e., the principal axis of the modified hexagonal CEF could be fluctuating among the three directions, maintaining the almost hexagonal symmetry, before fixed in phase II below $T_{\text{N}}^{\;\prime\prime}$. 
The strong ultrasonic attenuation observed in the $C_{66}$ mode in phase I could be indicative of such an unstable state of the lattice. 
The single peak of the $(6,0,0)$ reflection with a decrease in the $d$ spacing due to the in-plane AFM order also support such interpretation. 
We show in Fig.~\ref{fig:SpinFlop}(c) an illustration of the Dy-$4f$ charge distribution in this phase, representing small in-plane magnetic anisotropy though with small amount of $O_{22}$ and $O_{xy}$ quadrupole moments. 

As the magnitude of the ordered moment increases with decreasing $T$, the C1+C2 phase becomes more favored than the IC phase due to the increase in the CEF anisotropy. 
Then, the intensity of the IC peak begins to decrease. 
The commensurate $q=0.5$ can be interpreted as a compromise between the RKKY-interaction $J(\bm{q})$ and the hexagonal CEF anisotropy. 
Although the DM interaction should favor the single helicity of the C1 helimagnetic structure, the opposite helicity is mixed in the actual ordered state. This suggests that the DM interaction plays a less important role in the C1+C2 phase than in the IC phase; the RKKY interaction along the $c$-axis and the hexagonal CEF anisotropy govern the magnetic order, although both are much weaker than the in-plane AFM interaction. 
The coexistence of the IC phase and the C1+C2 phase, with pure helicity and mixed helicity, respectively, reflects a competing nature of the DM exchange interaction and the hexagonal CEF anisotropy.  

Finally, on entering phase II below $T_{\text{N}}^{\;\prime\prime}$, the FQ interactions become dominant in association with the development of the ordered magnetic moments. The RKKY interaction, which prefers the incommensurate order, becomes less important.
This results in a fixed monoclinic lattice distortion with different CEF anisotropy at the Dy-1 and Dy-2 sites. 
This causes the spin-flop transition by temperature as a result of the competition between the Zeeman energy and the anisotropy energy as discussed in the previous subsection. At the lowest temperature, the ordered structure is governed by the strong CEF anisotropy and the strong in-plane AFM interaction. 

\section{Conclusion}
\label{sec:V}
We have studied the successive phase transitions in a chiral magnet DyNi$_3$Ga$_9$, in which a FQ ordering and a helimagnetic ordering have been suggested to coexist below the N\'{e}el ordering temperature at $T_{\text{N}}=10$ K.  
The incommensurate (IC) helimagnetic order first develops just below $T_{\text{N}}$ with $\bm{q}=(0, 0, q)$, where $q\sim 0.43$ reflects the RKKY interaction along the $c$ axis, followed by the first order transition at $T_{\text{N}}^{\;\prime}=9.0$ K to the commensurate (C) helimagnetic order with $q=0.5$ and 1.5. 
In the IC-helimagnetic structure, the strongly coupled AFM moments of Dy in the honeycomb layer rotate counterclockwise on propagating along the $c$-axis in the right handed crystal, indicating that the chiral degeneracy is lifted by the much weaker DM interaction. 
Although the IC-helimagnetic structure has a pure helicity, the C-helimagnetic structure has a mixed helicity. 
This result shows that the C-helimagnetic structure is more influenced by the crystal field anisotropy than the DM interaction. 
Below $T_{\text{N}}^{\;\prime\prime}=8.5$ K, the helimagnetic peaks completely disappear and the canted antiferromagnetic order with $\bm{q}=(0, 0, 0)$ develops, accompanied by a lattice distortion which is inferred to be monoclinic.  
The canted-AFM structure exhibits a spin-flop transition at 5 K by sweeping temperature in a weak magnetic field. 
All these features are caused by the competition among the DM interaction, RKKY interaction, crystal field anisotropy, and the Zeeman energy in a dominating in-plane AFM interaction. 

\acknowledgments
The authors acknowledge valuable discussions with I. Ishii and S. Nakamura. 
This work was supported by JSPS KAKENHI Grant Numbers JP18K18737 and JP20H01854. 
This work was also supported by Chirality Research Center (Crescent) in Hiroshima University and JSPS Core-to-Core Program, A. Advanced Research Networks. 
MT is supported by the Hiroshima University Graduate School Research Fellowship. 
The synchrotron radiation experiments at SPring-8 were performed under Proposal No. 2019A3711 (BL22XU). 
The synchrotron radiation experiments at KEK was performed under the approval of the Photon Factory Program Advisory Committee (Proposal No. 2020G034). 
A part of this work was supported by QST Advanced Characterization Nanotechnology Platform under the remit of ``Nanotechnology Platform'' of the Ministry of Education, Culture, Sports, Science and Technology (MEXT), Japan (Grant No. JPMXP09A19QS0020). 
\appendix

\section{Chirality of the crystal}
\begin{figure}[t]
\begin{center}
\includegraphics[width=8cm]{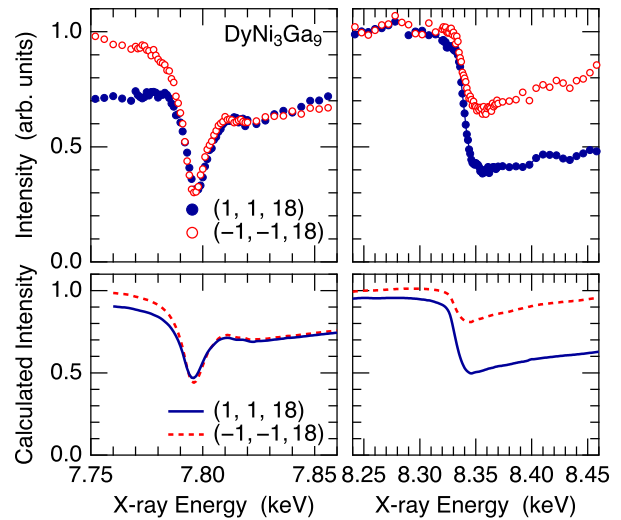}
\end{center}
\caption{top: X-ray energy dependences of the (1, 1, 18) and ($\bar{1}$, $\bar{1}$, 18) fundamental reflections around the Dy $L_3$-edge and the Ni $K$-edge. bottom: Calculated intensities of the (1, 1, 18) and ($\bar{1}$, $\bar{1}$, 18) reflections assuming a right-handed crystal. }
\label{fig:Escan1118}
\end{figure}
When the atomic position of Ni at the $18f$ Wycoff position in the $R32$ space group is written by $(x, y, z)=(0.3335, 0.0060, 0.08452)$~\cite{Ninomiya17,Gladyshevskii93}, we call the crystal as right handed. 
The Ni sites of the mirror reflected crystal is expressed by $(x, y, z)=(0.3335, 0.3275, 0.08452)$, which we call as left handed. 
We can distinguish the crystal chirality of the sample used in the present experiment, especially at the spot where the x-ray beam is irradiated, by measuring the energy dependences of the two Bragg reflections of a Bijovet pair. 
When the structure factor of the $hkl$ reflection for a right-handed crystal is expressed by 
$F_{\text{R},hkl}(\omega)=c_{\text{Dy}}f_{\text{Dy}}(\omega)+c_{\text{Ni}}f_{\text{Ni}}(\omega)+c_{\text{Ga}}f_{\text{Ga}}(\omega)$, 
$F_{\text{R},\bar{h}\bar{k}\bar{l}}(\omega)$ is given by taking the complex conjugates of the coefficients $c_{\text{Ni}}$ and $c_{\text{Ga}}$. 
Note that $c_{\text{Dy}}$ is real because the Dy atoms are located at $(0,0,z)$ and $(0,0,-z)$. 
The structure factor of a $h'k'l'$ reflection related by a proper symmetry operation (rotation) to $\bar{h}\bar{k}\bar{l}$, a Bijovet pair of $hkl$, is equal to $F_{\text{R},\bar{h}\bar{k}\bar{l}}(\omega)$. Then, if $c_{\text{Ni}}$ and $c_{\text{Ga}}$ are complex, $|F_{\text{R},hkl}(\omega)|$ and $|F_{\text{R},h'k'l'}(\omega)|$ are different, leading to different intensities which are more enhanced near the absorption edges. 
In the present experiment, we measured the intensities of a Bijovet pair reflections of $(1,1,18)$ and $(\bar{1},\bar{1},18)$. 

Figure~\ref{fig:Escan1118} shows the x-ray energy dependences of the intensity measured around the absorption edges of Dy and Ni. 
Shown in the bottom columns are the calculated intensities assuming a right-handed crystal. The imaginary part of the anomalous scattering factor was obtained from the fluorescence spectrum measured around the edges and the real part was obtained from the Kramers-Kronig transformation. 
We used calculated anomalous scattering factors at energies far from the absorption edges~\cite{Sasaki89}. 
If the crystal structure is left-handed, the relation $F_{\text{L},hkl}(\omega)=F_{\text{R},h'k'l'}(\omega)$ holds. 
Therefore, if the calculated intensities for  $(1,1,18)$ and $(\bar{1},\bar{1},18)$ have opposite relations to those the observations, we should conclude that the crystal is left-handed. 
Since the observations and the calculations are consistent, we can conclude that the crystal chirality of the sample used in this experiment is right-handed.

\section{Temperature hysteresis of the helimagnetic orderings}
\begin{figure}[t]
\begin{center}
\includegraphics[width=7.5cm]{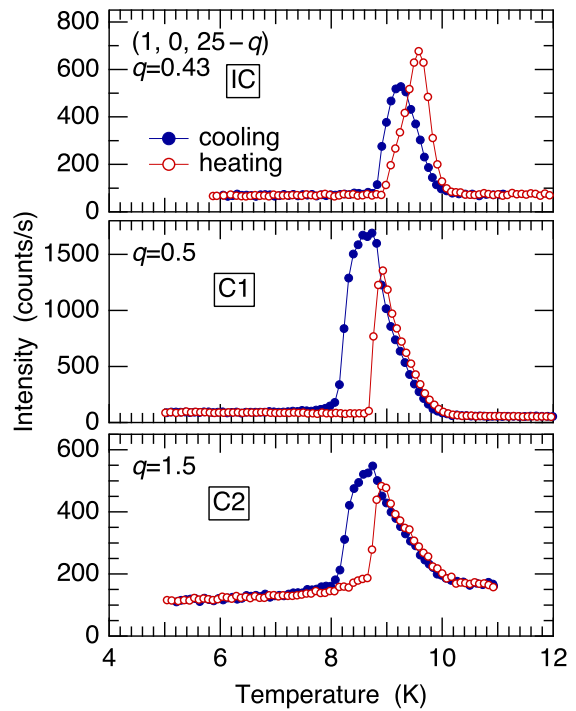}
\end{center}
\caption{Temperature dependences of the peak intensity of the $(1, 0, 25-q)$ reflections for $q=0.43$ (IC), 0.5 (C1), and 1.5 (C2), at the $E1$ resonance energy of 7.794 keV without polarization analysis. The intensity of the IC peak was measured at a fixed position of $q=0.43$. 
The temperature was swept at a rate of 0.4 K/min. }
\label{fig:Thysteresis}
\end{figure}
Figure~\ref{fig:Thysteresis} shows the temperature hysteresis of the peak intensities between $T_{\text{N}}^{\;\prime\prime}$ and $T_{\text{N}}$ measured by sweeping the temperature at a fixed detector position. 
Note that the detector is fixed at $q=0.43$ in the measurement of the IC peak, not following the the peak top intensity precisely. 
It is also noted that the temperature is swept at a rate of 0.4 K/min, not reflecting the accurate temperature of the sample itself. 
The purpose of this figure is only to show the temperature hysteresis. 

With decreasing temperature, as described in the main text, the IC peak first develops when crossing $T_{\text{N}}=10$ K. 
The C1 and C2 peaks develop next, and they are more enhanced when the temperature crosses the boundary at $T_{\text{N}}^{\;\prime}$, where the IC peak starts to diminish. Then, they disappear abruptly at $T_{\text{N}}^{\;\prime\prime}$, which is lower than that in the heating process. 
In the heating process, these transitions shift to higher temperatures, indicating that these are first order transitions. 
In both heating and cooling processes, the C1 and C2 peaks are proportional to each other, indicating that these peaks are the Fourier components of the identical magnetic structure.  


\clearpage
\begin{center}
\textbf{\large{Supplemental Material}}
\end{center}
\vspace{5mm}

\renewcommand{\topfraction}{1.0}
\renewcommand{\bottomfraction}{1.0}
\renewcommand{\dbltopfraction}{1.0}
\renewcommand{\textfraction}{0.01}
\renewcommand{\floatpagefraction}{1.0}
\renewcommand{\dblfloatpagefraction}{1.0}
\setcounter{topnumber}{5}
\setcounter{bottomnumber}{5}
\setcounter{totalnumber}{10}

\renewcommand{\theequation}{S\arabic{equation}}
\renewcommand{\thefigure}{S\arabic{figure}}
\renewcommand{\thetable}{S-\Roman{table}}
\setcounter{section}{19}
\setcounter{figure}{0}

\section*{splitting of the 600 peak and the lattice distortion}
\begin{figure*}[t]
   \begin{minipage}[b]{0.47\textwidth}
       \begin{center}
       \includegraphics*[width=7.5cm]{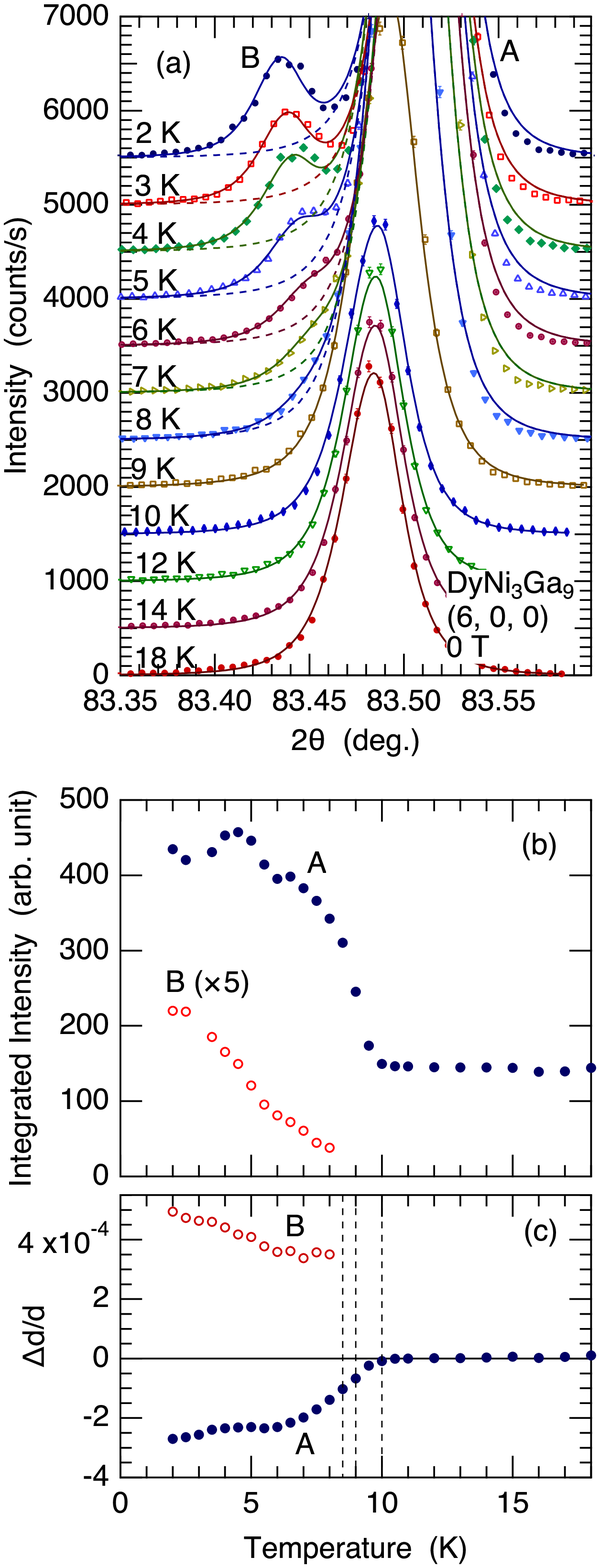}
       \end{center}
\caption{(a) Expanded view of the $\theta$-$2\theta$ scans for Fig.~12 in the main text. 
(b) Temperature dependence of the integrated intensity of the peak A and B. The intensity for the peak B is multiplied by five. 
(c) Temperature dependence of $\Delta d/d$, the same as Fig.~12(b) in the main text.  }
\label{fig:th2th600}
   \end{minipage}
\hfill
   \begin{minipage}[b]{0.47\textwidth}
       \begin{center}
       \includegraphics*[width=8cm]{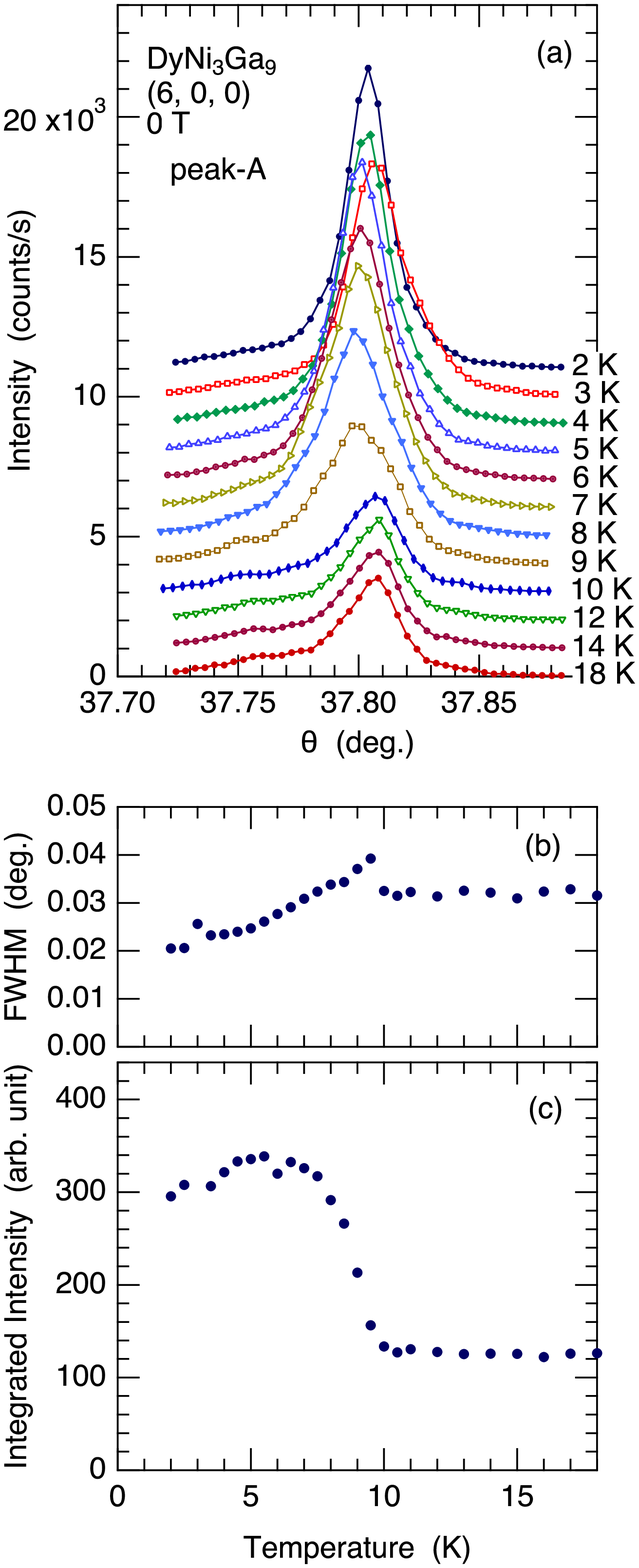}
       \end{center}
\caption{(a) Temperature dependence of the rocking-scan for the peak-A in the $\theta$-$2\theta$ scan. 
(b) Temperature depdnece of the full-width at half-maximum deduced from the fits with Lorentzian functions. 
(c) Temperature dependence of the integrated intensity for the rocking-scans in (a).  }
\label{fig:thscan600}
   \end{minipage}
\end{figure*}

Although we have successfully detected the peak splitting of the 600 fundamental reflection in the canted-AFM phase below $T_{\text{N}}^{\;\prime\prime}$, the lattice distortion in the intermediate phase between $T_{\text{N}}^{\;\prime\prime}$ and $T_{\text{N}}$ remains an open question. 
Figure \ref{fig:th2th600}(a) shows the same data for Fig.~12 in the main text in an expanded scale. 
The temperature dependences of the integrated intensity for the peak A and B are shown in Fig. \ref{fig:th2th600}(b). 
As described in the main text, the peak split is clearly visible below $T_{\text{N}}^{\;\prime\prime}$. 
However, it is unclear above $T_{\text{N}}^{\;\prime\prime}$ whether the splitting vanishes or it is too small to be resolved and continuously decreases to zero toward $T_{\text{N}}$. 
There is also a clear anomaly at $T_{\text{N}}$, below which the peak intensity increases and the peak position begins to shift. 
This shows that some kind of lattice distortion occurs below $T_{\text{N}}$. 

With respect to the increase in intensity below $T_{\text{N}}$, it is probably due to the reduction of the extinction effect by the occurrence of lattice distortion. The crystal mosaicity would increase and therefore the extinction decreases, leading to the increase in the peak intensity. 
However, the expected increase in the crystal mosaicity does not seem to be reflected in the rocking scan. 
In Fig.~\ref{fig:thscan600}(a), we show the temperature dependence of the rocking scans for the peak A in the $\theta$-$2\theta$ scans in Fig.~\ref{fig:th2th600}(a). The temperature dependence of the full width at half maximum (FWHM) and the integrated intensity is shown in Fig.~\ref{fig:thscan600}(b) and (c), respectively. 
Although the peak width exhibits a small increase just below $T_{\text{N}}$, the FWHM decreases with decreasing temperature. 
Since we have no experimental access to observe the microscopic mechanism of the extinction, we only speculate that the lattice distortion in a microscopic scale reduces the extinction effect, which is not reflected in the rocking curve and the width of the Bragg peak.

\end{document}